\documentclass[a4paper,english]{lipics-v2018}

\usepackage{microtype}

\newcommand{\set}[1]{\ensuremath{\{ #1 \}}}
\newcommand{\abs}[1]{\ensuremath{\lvert #1 \rvert}}
\newcommand{\Oh}[1]{\ensuremath{\mathsf{O}\!\left( #1 \right)}}

\newcommand{\kmer}[1]{$#1$\nobreakdash-mer}

\newcommand{\greverse}[1]{\ensuremath{\overline{#1}}}

\newcommand{\rank}{\ensuremath{\mathsf{rank}}}
\newcommand{\select}{\ensuremath{\mathsf{select}}}
\newcommand{\LF}{\ensuremath{\mathsf{LF}}}
\newcommand{\find}{\ensuremath{\mathsf{find}}}
\newcommand{\locate}{\ensuremath{\mathsf{locate}}}
\newcommand{\extract}{\ensuremath{\mathsf{extract}}}

\newcommand{\SA}{\ensuremath{\mathsf{SA}}}
\newcommand{\DA}{\ensuremath{\mathsf{DA}}}
\newcommand{\BWT}{\ensuremath{\mathsf{BWT}}}
\newcommand{\Carray}{\ensuremath{\mathsf{C}}}
\newcommand{\pruned}[1]{\textsf{pruned-$#1$}}
\newcommand{\unfolded}[1]{\textsf{unfolded-$#1$}}

\title{Haplotype-aware graph indexes}

\titlerunning{Haplotype-aware graph indexes}

\author{Jouni Sirén}{University of California, Santa Cruz, USA \\ Wellcome Sanger Institute, Hinxton, UK}%
{jouni.siren@iki.fi}{https://orcid.org/0000-0001-5828-4139}{Funded by Wellcome Trust grant WT206194, the National Institutes of Health (5U41HG007234, 1U01HL137183-01), and the W. M. Keck Foundation (DT06172015)}

\author{Erik Garrison}{Wellcome Sanger Institute, Hinxton, UK}%
{eg10@sanger.ac.uk}{https://orcid.org/0000-0003-3821-631X}{Funded by Wellcome Trust grant WT206194.}

\author{Adam M. Novak}{University of California, Santa Cruz, USA}%
{anovak@soe.ucsc.edu}{https://orcid.org/0000-0001-5828-047X}{Funded by the National Institutes of Health (5U41HG007234, 1U01HL137183-01) and the W. M. Keck Foundation (DT06172015)}

\author{Benedict Paten}{University of California, Santa Cruz, USA}%
{bpaten@ucsc.edu}{https://orcid.org/0000-0001-8863-3539}{Funded by the National Institutes of Health (5U41HG007234, 1U01HL137183-01) and the W. M. Keck Foundation (DT06172015)}

\author{Richard Durbin}{Department of Genetics, University of Cambridge, UK \\ Wellcome Sanger Institute, Hinxton, UK}%
{rd@sanger.ac.uk}{https://orcid.org/0000-0002-9130-1006}{Funded by Wellcome Trust grant WT206194.}

\authorrunning{J. Sirén, E. Garrison, A.\,M. Novak, B. Paten, and R. Durbin}

\Copyright{Jouni Sirén, Erik Garrison, Adam M. Novak, Benedict Paten, and Richard Durbin}

\subjclass{\ccsdesc[500]{Theory of computation~Pattern matching}, \ccsdesc[300]{Theory of computation~Data compression}, \ccsdesc[500]{Applied computing~Computational genomics}}

\keywords{FM-indexes, variation graphs, haplotypes}

\supplement{The implementation can be found at \url{https://github.com/vgteam/vg}, \url{https://github.com/jltsiren/gbwt}, and \url{https://github.com/jltsiren/gcsa2}.}

\EventEditors{Laxmi Parida and Esko Ukkonen}
\EventNoEds{2}
\EventLongTitle{18th International Workshop on Algorithms in Bioinformatics (WABI 2018)}
\EventShortTitle{WABI 2018}
\EventAcronym{WABI}
\EventYear{2018}
\EventDate{August 20--22, 2018}
\EventLocation{Helsinki, Finland}
\EventLogo{}
\SeriesVolume{113}
\ArticleNo{4}
\nolinenumbers 
\hideLIPIcs  

\begin{document}

\maketitle

\begin{abstract}
The variation graph toolkit (VG) represents genetic variation as a graph. Each path in the graph is a potential haplotype, though most paths are unlikely recombinations of true haplotypes. We augment the VG model with haplotype information to identify which paths are more likely to be correct. For this purpose, we develop a scalable implementation of the graph extension of the positional Burrows--Wheeler transform. We demonstrate the scalability of the new implementation by indexing the 1000 Genomes Project haplotypes. We also develop an algorithm for simplifying variation graphs for k-mer indexing without losing any k-mers in the haplotypes.
\end{abstract}

\section{Introduction}

Sequence analysis pipelines often start by mapping the sequence reads to a \emph{reference genome} of the same species. A read aligner first uses a \emph{text index} to find candidate positions for the read. Then it aligns the read to the candidate positions, trying to find the best mapping.

The reference genome may represent the sequenced individual poorly, e.g.\ when it is a single sequence that diverges substantially at some location. Mapping reads to such a reference can introduce \emph{reference bias} to the subsequent analysis. Richer reference models can help to avoid the bias, but challenges remain in choosing the right model and working with it effectively \cite{Marschall2018}.

We can replace the single reference sequence with a \emph{collection of haplotypes}. Because individual genomes are similar, compressed text indexes can store such collections in very little space \cite{Maekinen2010}. However, due to this similarity, most reads map equally well to many haplotypes. If the reference model is a simple collection, we cannot tell whether a read maps to the same position in different haplotypes or not.

If the haplotypes are \emph{aligned}, we can use the alignment to determine whether the mappings are equivalent. Text indexes can also take advantage of the alignment by storing shared substrings only once \cite{Huang2010}. The FM-index of alignment \cite{Na2016,Na2018} goes one step further by collapsing the multiple alignment into a \emph{directed acyclic graph} (DAG), where each node is labeled by a sequence. It indexes the graph and stores some additional information for determining which paths correspond to valid haplotypes.

We can also build a reference graph directly from a reference sequence and a set of \emph{variants} \cite{Schneeberger2009}. This approach has been used in many tools such as GCSA \cite{Siren2014}, BWBBLE \cite{Huang2013}, vBWT \cite{Maciuca2016}, the Seven Bridges Graph Pipeline \cite{Rakocevic2017}, and Graphtyper \cite{Eggertsson2017}. Algorithms for working with sequences are often easy to generalize to DAGs. On the other hand, because an acyclic graph imposes a global alignment on the haplotypes, allowing only matches, mismatches, and indels, it cannot represent structural variation such as duplications or inversions adequately.

Assembly graphs such as \emph{de~Bruijn graphs} collapse sequences by local similarity instead of global alignment. They are better suited to handling structural variation than DAGs. However, the lack of a global coordinate system limits their usefulness as references.

Graph-based reference models share certain weaknesses. Because they collapse sequences between variants, they represent both the original haplotypes and their \emph{recombinations}, that is paths that switch between haplotypes. This may cause false positives when a read maps better to an unobserved recombination than to the correct path. Graph regions with many variants in close proximity can give rise to very large numbers of recombinant paths, and be \emph{too complex} to index completely. Graph tools try to deal with such regions by, for example, limiting the amount of variation in the graph, artificially simplifying complex regions, and making trade-offs between query performance, index size, and maximum query length.

CHOP \cite{Mokveld2018} embeds haplotypes into a graph and indexes the corresponding paths. For a given parameter $k$, the graph is transformed into a collection of short strings such that adjacent strings overlap by $k-1$ characters. Each haplotype can be represented as a sequence of adjacent strings. Any read aligner can be used to map reads to the strings. However, because the aligner sees only short strings, it cannot map long reads or paired-end reads.

The \emph{variation graph toolkit} (VG) \cite{Garrison2017} works with many kinds of graphs. While the tools mentioned earlier use graphs to represent other information (e.g.\ sequences or variants), the graph itself is the primary object in the VG model. A global coordinate system can be provided by designating certain paths as \emph{reference paths}.

VG uses GCSA2 \cite{Siren2017} as its text index. GCSA2 represents a \kmer{k} index as a de~Bruijn graph and compresses it structurally by merging redundant nodes. VG handles complex graph regions by indexing a \emph{simplified graph}, although the final alignment is done in the original graph. The drawback of this approach is that simplification can break paths corresponding to known haplotypes, while leaving paths representing recombinations intact.

In this paper, we augment the VG model with haplotype information. We develop the GBWT, a scalable implementation of the graph extension of the positional Burrows--Wheeler transform (gPBWT) \cite{Durbin2014,Novak2017a}, to store the haplotypes as paths in the graph. To demonstrate the scalability of the GBWT, we build an index for the the 1000 Genomes Project haplotypes. We also describe an algorithm that adds the haplotype paths back to the simplified graph, without reintroducing too much complexity.

The main differences to the old gPBWT implementation \cite{Novak2017a} are:
\begin{itemize}
\item We use \emph{local structures} for each node instead of global structures for the graph. The index is smaller and faster and takes better advantage of memory locality.

\item The GBWT is implemented as an ordinary text index instead of a special-purpose index for paths. Most FM-index algorithms can be used with it. For example, we can use the GBWT as an FMD-index \cite{Li2012} and support \emph{bidirectional search}.

\item We have a fast and space-efficient \emph{incremental construction} algorithm that does not need access to the entire collection of haplotypes at the same time.

\item Our implementation can be used \emph{independently} of VG.
\end{itemize}

The haplotype information stored in GBWT can also be used to improve read mapping, and potentially, variant inference. We leave this investigation to a subsequent paper.

\section{Background}

\subsection{Strings and graphs}

A \emph{string} $S[0, n-1] = s_{0} \dotsm s_{n-1}$ of length $\abs{S} = n$ is a sequence of \emph{characters} over an \emph{alphabet} $\Sigma = \set{0, \dotsc, \sigma-1}$. \emph{Text} strings $T[0,n-1]$ are terminated by an \emph{endmarker} $T[n-1] = \$ = 0$ that does not occur anywhere else in the text. \emph{Substrings} of string $S$ are sequences of the form $S[i, j] = s_{i} \dotsm s_{j}$. We call substrings of length $k$ \kmer{k}s and substrings of the type $S[0, j]$ and $S[i, n-1]$ \emph{prefixes} and \emph{suffixes}, respectively.

Let $S[0, n-1]$ be a string. We define $S.\rank(i, c)$ as the number of occurrences of character $c$ in the prefix $S[0, i-1]$. We also define $S.\select(i, c) = \max \set{j \le n \mid S.\rank(j, c) < i}$ as the position of the occurrence of rank $i > 0$. A \emph{bitvector} is a data structure that stores a binary sequence and supports efficient $\rank$/$\select$ queries over it.

A \emph{graph} $G = (V, E)$ consists of a finite set of \emph{nodes} $V \subset \mathbb{N}$ and a set of \emph{edges} $E \subseteq V \times V$. We assume that the edges are \emph{directed}: $(u, v) \in E$ is an edge \emph{from} node $u$ \emph{to} node $v$. The \emph{indegree} of node $v$ is the number of \emph{incoming} edges to $v$, while the \emph{outdegree} is the number of \emph{outgoing} edges from $v$. Let $P = v_{0} \dotsm v_{\abs{P}-1}$ be a string over the set of nodes $V$. We say that $P$ is a \emph{path} in graph $G = (V, E)$, if $(v_{i}, v_{i+1}) \in E$ for all $0 \le i < \abs{P} - 1$.

The VG model \cite{Garrison2017} is based on \emph{bidirected} graphs, where each node has two \emph{orientations}. We simulate them with directed graphs. We partition the set of nodes $V$ into \emph{forward} nodes $V_{f}$ and \emph{reverse} nodes $V_{r}$, with $V_{f} \cap V_{r} = \emptyset$ and $\abs{V_{f}} = \abs{V_{r}}$. We match each forward node $v \in V_{f}$ with the corresponding reverse node $\greverse{v} \in V_{r}$, with $\greverse{\greverse{v}} = v$ for all $v \in V_{f}$. For all nodes $u, v \in V$, we also require that $(u, v) \in E \iff (\greverse{v}, \greverse{u}) \in E$.

\subsection{FM-index}\label{sect:fm-index}

The \emph{suffix array} $\SA[0, n-1]$ of text $T[0, n-1]$ is an array of pointers to the suffixes of the text in \emph{lexicographic order}. For all $i < j$, we have $T[\SA[i], n-1] < T[\SA[j], n-1]$. The \emph{Burrows--Wheeler transform} (BWT) \cite{Burrows1994} is a permutation of the text with a similar combinatorial structure. We define it as string $\BWT[0, n-1]$, where $\BWT[i] = T[(\SA[i]-1) \bmod n]$. Let $\Carray[c]$ be the number of occurrences of characters $c' < c$ in the text. The main operation in BWT is the \emph{LF-mapping}, which we define as $\LF(i, c) = \Carray[c] + \BWT.\rank(i, c)$. We use shorthand $\LF(i)$ for $\LF(i, \BWT[i])$ and note that $\SA[\LF(i)] = (\SA[i] - 1) \bmod n$

Let $X$ be a string and let $c$ be a character. If $T' < X$ for $i$ suffixes $T'$ of text $T$, we say that string $X$ has \emph{lexicographic rank} $i$ among the suffixes of text $T$. The number of suffixes starting with any character $c' < c$ is $\Carray[c]$, and the number of suffixes $T' < X$ preceded by character $c$ is $\BWT.\rank(i, c)$. Hence the lexicographic rank of string $cX$ is $\LF(i, c)$.

The FM-index \cite{Ferragina2005a} is a text index based on the BWT. Assume that we can compute $\BWT.\rank(i, c)$ in $t_{r}$ time. Further assume that we have stored $(i, \SA[i])$ for all $\SA[i]$ divisible by some integer $d > 0$. The FM-index supports the following queries:
\begin{itemize}

\item $\find(X)$: Return the \emph{lexicographic range} $[sp, ep]$ of suffixes starting with \emph{pattern} $X$. If $[sp_{i+1}, ep_{i+1}]$ is the lexicographic range for pattern $X[i+1, \abs{X}-1]$, the range for pattern $X[i, \abs{X}-1]$ is $[\LF(sp_{i+1}, X[i]), \LF(ep_{i+1}+1, X[i]) - 1]$. By extending the pattern backwards, we can support $\find(X)$ in $\Oh{\abs{X} \cdot t_{r}}$ time.

\item $\locate(sp, ep)$: Return the \emph{occurrences} $\SA[sp, ep]$. For each $i \in [sp, ep]$, we iterate $\LF(i)$ until we find a stored pair $(\LF^{k}(i), \SA[\LF^{k}(i)])$. Then $\SA[i] = \SA[\LF^{k}(i)] + k$. Locating each occurrence $\SA[i]$ takes $\Oh{d \cdot t_{r}}$ time.

\item $\extract(j, j')$: Return the substring $T[j, j']$. We start from the nearest stored $(i, \SA[i])$ with $\SA[i] > j'$ and iterate $(i, \SA[i]) \leftarrow (\LF(i), \SA[i] - 1)$ until $\SA[i] = j + 1$. As $\BWT[i] = T[\SA[i]-1]$, we extract the substring backwards in $\Oh{(d+j'-j) \cdot t_{r}}$ time.

\end{itemize}

We can generalize the FM-index to indexing \emph{multiple texts} $T_{0}, \dotsc, T_{m-1}$. Each text $T_{j}$ is terminated by a distinct endmarker $\$_{j}$, where $\$_{j} < \$_{j+1}$ for all $j$. As the suffixes of the texts are all distinct, we can sort them unambiguously. In the final BWT, we replace each $\$_{j}$ with $\$$ in order to reduce alphabet size. The index works as with a single text, except that we cannot compute $\LF(i, \$)$. We also define the \emph{document array} $\DA$ as an array of \emph{text identifiers}. If $\SA[i]$ points to a suffix of text $T_{j}$, we define $\DA[i] = j$.

\section{Indexing haplotypes}

\subsection{Positional BWT}

Assume that we have $m$ haplotype strings $S_{0}, \dotsc, S_{m-1}$ of equal length over alphabet $\Sigma$. At each variant site $i$, character $S_{j}[i]$ tells whether haplotype $j$ contains the reference allele ($S_{j}[i] = 0$) or an alternate allele ($S_{j}[i] > 0$). Given a pattern $X$ and a range of sites $[i, i']$, we want to find the haplotypes $S_{j}$ matching the pattern at the specified sites ($S_{j}[i, i'] = X$). Ordinary FM-indexes do not support such queries, as they find all occurrences of the pattern.

The \emph{positional BWT} (PBWT) \cite{Durbin2014} is an FM-index that supports \emph{positional queries}. We can interpret it as the FM-index of texts $T_{0}, \dotsc, T_{m-1}$ such that $T_{j}[i] = (i, S_{j}[i])$ \cite{Gagie2017a}. If we want to search for pattern $X$ in range $[i, i']$, we search for pattern $X' = (i, X[0]) \dotsm (i', X[\abs{X}-1])$ in the FM-index. The texts are over a large alphabet, but their \emph{first-order empirical entropy} is low. We can encode the BWT using alphabet $\Sigma$ with a simple model. Assume that $\SA[x]$ points to a suffix starting with $(i+1, c)$. We often know the character from a previous query, and we can determine it using the $\Carray$ array. Then $\BWT[x] = (i, c')$ for a $c' \in \Sigma$, and we can encode it as $c'$. (Note that we build the $\rank$ structure for the original BWT, not the encoded BWT.) Because the collection of haplotype strings is usually \emph{repetitive}, we can compress the PBWT further by \emph{run-length encoding} the BWT \cite{Maekinen2010}.

\subsection{Graph extension}

Haplotypes correspond to paths in the VG model. Because chromosome-length phasings are often not available, there may be multiple paths for each haplotype. The \emph{graph extension} of the PBWT \cite{Novak2017a} generalizes the PBWT to indexing such paths. While the original extension was specific to VG graphs, we present a simplified version over directed graphs. We call this structure the \emph{Graph BWT} (GBWT), as it encodes the BWT using the graph as a model.

Let $P_{0}, \dotsc, P_{m-1}$ be paths in graph $G = (V, E)$. We can interpret the paths as strings over alphabet $V$. Assume that $0 \not\in V$, as we use it as the endmarker. We build an FM-index for the \emph{reverse strings}. We sort reverse prefixes in lexicographic order, so the LF-mapping traverses edges in the correct direction, and place the endmarker before the string.

For each node $v \in V$, we define the \emph{local alphabet} $\Sigma_{v} = \set{w \in V \mid (v, w) \in E }$. We also add $\$$ to $\Sigma_{v}$ if $v$ is the last node on a path, and define $\Sigma_{\$}$ as the set of the initial nodes on each path. We partition the BWT into substrings $\BWT_{v}$ corresponding to the prefixes ending with $v$, and encode each substring $\BWT_{v}$ using the local alphabet $\Sigma_{v}$. If $w \in \Sigma_{v}$ is the $k$th character in the local alphabet in sorted order, we encode it as $\Sigma_{v}(w) = k$.

The GBWT supports the following variants of the standard FM-index queries:
\begin{itemize}

\item $\find(X)$ returns the lexicographic range of reverse prefixes starting with the reverse pattern (the range of prefixes ending with the pattern).

\item $\locate(sp, ep)$ returns the haplotype identifiers $\DA[sp, ep]$. We do not return text offsets, as the node corresponding to the range $[sp, ep]$ already provides similar information.

\item $\extract(j)$ returns the path $P_{j}$. We save memory by not supporting substring extraction.

\end{itemize}

These queries should be understood as examples of what we can support. Because the GBWT is an FM-index for multiple texts, most algorithms using an FM-index can be adapted to use the GBWT. For example, let $P = v_{0} \dotsm v_{\abs{P}-1}$ be a path. The \emph{reverse path} of $P$ is the path $\greverse{P} = \greverse{v_{\abs{P-1}}} \dots \greverse{v_{0}}$ traversing the reverse nodes in the reverse order. If we also index $\greverse{P}$ for every path $P$, the GBWT becomes an FMD-index \cite{Li2012} that supports \emph{bidirectional search}.

\subsection{Records}\label{sect:records}

We develop a GBWT representation based on the following assumptions:
\begin{enumerate}

\item Almost all nodes $v \in V$ have a low outdegree, making the local alphabet $\Sigma_{v}$ small. Hence we can afford storing the $\rank$ of all $w \in \Sigma_{v}$ at the start of $\BWT_{v}$. Decompressing that information every time we access the node does not take too much time either.

\item The number of occurrences of almost all nodes is bounded by the number of haplotypes. As the length of $\BWT_{v}$ is bounded for almost all $v \in V$, we can afford scanning it every time we compute $\rank$ within it.

\item The collection of paths is repetitive. Run-length encoding compresses the BWT well, reducing both index size and the time required for scanning $\BWT_{v}$.

\item The set of nodes $V$ is a dense subset of a range $[a, b]$. Hence we can afford storing some information for all $i \in [a, b]$ without using too much space.

\item The graph is almost linear and almost topologically sorted. The closer to topological order we can store the nodes, the less space we need for graph topology, and the better we can take advantage of \emph{memory locality}.

\end{enumerate}

\begin{figure}
\includegraphics[width=\textwidth]{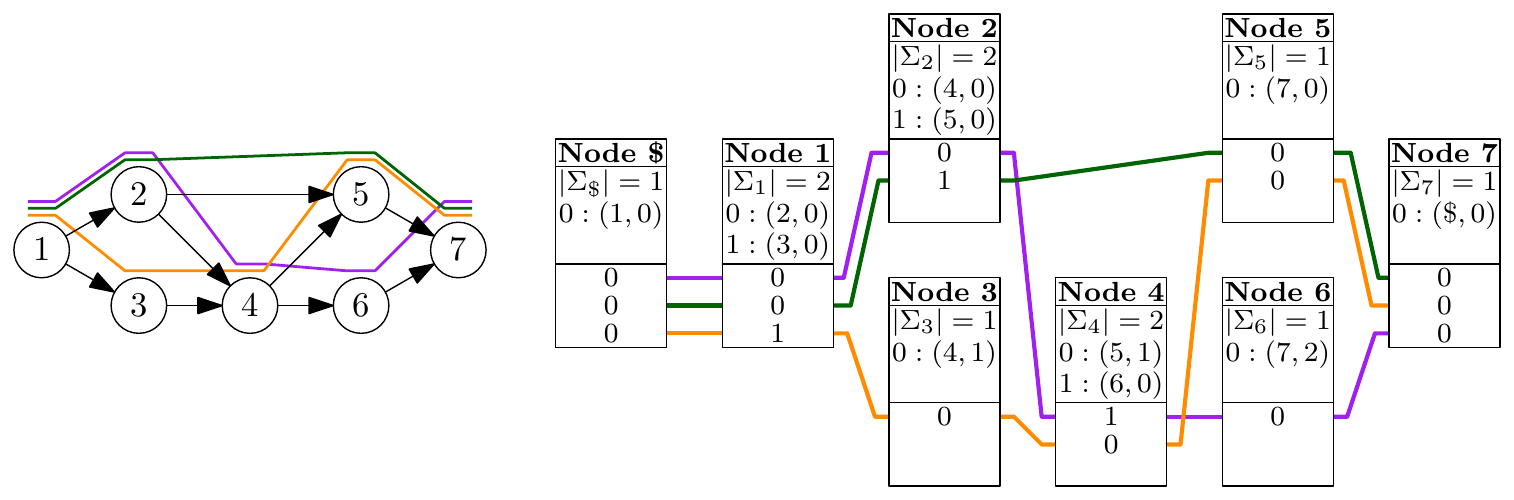}
\caption{Left: A graph with three paths. Right: GBWT of the paths.}\label{fig:gbwt-example}
\end{figure}

We store a \emph{record} consisting of a \emph{header} and a \emph{body} for each node $v \in V$ and for the endmarker $\$$. For each character $w \in \Sigma_{v}$ in sorted order, the header stores a pair $(w, \BWT.\rank(v, w))$, where $\BWT.\rank(v, w)$ is the total number of occurrences of character $w$ in all $\BWT_{v'}$ with $v' < v$. The body run-length encodes $\BWT_{v}$, representing a run of $\ell$ copies of character $w$ as a pair $(\Sigma_{v}(w), \ell)$. See Figure~\ref{fig:gbwt-example} for an example.

Because the BWT is a set of records, we use node/offset pairs as \emph{positions}. Pair $(v, i)$ refers to offset $\BWT[\Carray[v]+i] = \BWT_{v}[i]$. We define $\rank$ queries over positions as
$$
\BWT.\rank((v, i), w) = \BWT.\rank(v, w) + \BWT_{v}.\rank(i, w).
$$
Similarly, we define $\LF((v, i), w) = (w, \BWT.\rank((v, i), w))$ and use it in place of ordinary LF-mapping in the FM-index.

The FM-index is based on iterating LF-mapping. Because LF-mapping tends to jump randomly around the BWT, this can be a significant bottleneck. GBWT achieves better memory locality, if we store the records for adjacent nodes close to each other. When we iterate LF-mapping over a path in the graph, we traverse adjacent memory regions.

As a run-length encoded FM-index, GBWT supports the \emph{fast} $\locate()$ algorithm \cite{Maekinen2010}. The \emph{direct algorithm}, as described in Section~\ref{sect:fm-index}, locates each position $i \in [sp, ep]$ separately. If we instead process the entire range at once, advancing every position by one step of LF-mapping at the same time, we achieve better memory locality. We can also compute LF-mapping for an entire run $\BWT_{v}[x, y] = w^{y+1-x}$ in the same time as for a single position $i \in [x, y]$.

\subsection{GBWT encodings}\label{sect:encodings}

\emph{Dynamic GBWT} is intended for index construction, where speed is more important than size. We have an array of fixed-size records for characters $\$$ and $v \in [a, b]$, including character values $v \not\in V$. The record for $v$ has four pointers to arrays: header, body, incoming edges, and haplotype identifiers. For each incoming edge $(u, v) \in E$, the incoming edges array stores a pair $(u, \BWT_{u}.\rank(\abs{\BWT_{u}}, v))$, recording the number of paths crossing from $u$ to $v$.

Let $\SA_{v}$ and $\DA_{v}$ be the parts of $\SA$ and $\DA$ corresponding to $\BWT_{v}$. The haplotype identifiers array for node $v$ stores, in sorted order, pairs $(i, \DA_{v}[i])$ for which $\SA_{v}[i]$ points to either the last node on a path or a path position divisible by $d > 0$. These pairs are used for $\locate()$ queries, like stored $\SA$ pointers in an ordinary FM-index.

\emph{Compressed GBWT} balances query performance with index size. We use it when the set of haplotypes is fixed and for storing the index on disk. Each record is a byte array. We encode integers as sequences of bytes, where the lower 7~bits contain data and the high bit tells whether the encoding continues. The header starts with $\abs{\Sigma_{v}}$. We encode the outgoing edges $(w_{i}, \BWT.\rank(v, w_{i}))$ \emph{differentially}, replacing $w_{i}$ with $w_{i}-w_{i-1}$. If the local alphabet is large, each run $(k, \ell)$ in the body is encoded as an integer pair. Otherwise we encode $k$ and as much of $\ell$ as possible in the first byte, and continue with the remaining run length in subsequent bytes. We concatenate all records and mark their starting positions in a sparse bitvector \cite{Okanohara2007}. The records can be accessed with $\select$ queries on the bitvector.

Each compressed record must be decompressed sequentially. As the stored haplotype identifiers tend to cluster in certain nodes, storing them in records would make these records large and slow to decompress. Instead, we use a global structure for the haplotype identifiers. The structure consists of three bitvectors and an array of identifiers:
\begin{itemize}

\item Uncompressed bitvector $B_{s}$ marks the records with stored identifiers. If the $i$th record contains identifiers, we set $B_{s}[i] = 1$. This allows us to skip checking the identifiers in most records when iterating $\LF()$.

\item Sparse bitvector $B_{r}$ is defined over the concatenated offset ranges of the records with stored identifiers. If $B_{s}[i] = 1$, the range for the record starts at $B_{r}.\select(B_{s}.\rank(i, 1) + 1, 1)$.

\item Sparse bitvector $B_{o}$ covers the same range as $B_{r}$. If $B_{o}[i+j] = 1$ and the range for the record starts at $B_{o}[i]$, we have an identifier for offset $j$ at array position $B_{o}.\rank(i+j, 1)$.

\end{itemize}

\section{GBWT construction}

The assumptions in Section~\ref{sect:records} make the GBWT easier to build than an ordinary FM-index. \emph{Inserting} new texts into the collection updates adjacent records, just like searching traverses adjacent records. Because the local alphabet is small, because the number of occurrences of each character is limited, and because run-length encoding compresses the BWT well, records tend to be small. Hence we can afford \emph{rebuilding} a record each time we update it.

On the other hand, the GBWT is harder to build than the PBWT. In the PBWT, all strings are of the same length and have the same variant site at the same position. Hence we can build the final record for a site in a single step. In the GBWT, indels in the haplotypes become indels on the haplotype paths, and hence we have to update the same record multiple times. We also have to buffer the strings instead of indexing them as we generate them.

\subsection{Basic construction}

The following algorithm \cite{Hon2007} updates the BWT of text $T$ to be the BWT of text $cT$, where $c$ is a character. It forms the basis of many incremental BWT construction algorithms.
\begin{enumerate}

\item Find the offset $i$ where $\BWT[i] = \$$ and replace the endmarker with character $c$.

\item Compute $i' = \LF(i, c)$ and insert a new endmarker between offsets $i'-1$ and $i'$.

\end{enumerate}
If we have a BWT for $m$ texts, we can insert a new empty text by inserting an endmarker between offsets $m-1$ and $m$. By iterating the above algorithm, we can then insert the actual text. If we have a \emph{dynamic FM-index} \cite{Chan2007}, this can be quite efficient in practice.

The BCR algorithm \cite{Bauer2013} builds BWT for $m$ texts. It starts with the BWT for $m$ empty texts and then extends each text by one character in each step. Originally intended for indexing short reads, the BCR algorithm is also used for PBWT construction.

Our GBWT construction algorithm is similar to RopeBWT2 \cite{Li2014a}. We have a dynamic GBWT and insert multiple texts into the index in a single \emph{batch} using the BCR algorithm. In each step, we extend each text by one character. In the following, $v$ and $w$ are the current and the next character in the current text $T_{j}$ and $i$ is a record offset. If $v$ is the last character of the text (the endmarker is at $T_{j}[0]$), we set $w = \$$. In each step, we:
\begin{enumerate}

\item \textbf{Rebuild records:} The texts are sorted by positions $(v, i)$ such that the endmarker of that text should be at $\BWT_{v}[i]$. (We do not write the temporary endmarkers to the records.) We process all texts at the same node $v$ to rebuild the record.
\begin{enumerate}

\item If the record does not contain the edge $(v, w)$, we add $(w, 0)$ to the header.

\item We add BWT runs and haplotype identifiers until offset $i$ to the new record. If we have inserted $k$ characters so far, we replace haplotype identifier $(i', j')$ with $(i'+k, j')$.

\item If $w = \$$ or the text position is divisible by $d$, we insert haplotype identifier $(i, j)$.

\item We insert $w$ to the BWT and set $i \leftarrow \BWT_{v}.\rank(i, w)$.

\item If $w \ne \$$, we increment the number of paths from $v$ to $w$ in the incoming edges of $w$.

\end{enumerate}

\item \textbf{Sort:} We sort the texts by $(w, v, i)$, which is the order we need in the next step. If $w = \$$, the text is now fully inserted, and we remove it from further processing.

\item \textbf{Rebuild offsets:} For each distinct node $w$, we rebuild the $\BWT.\rank(v', w)$ fields in the outgoing edges of predecessor nodes $v'$ using the path counts in the incoming edges of $w$. Then we set $i \leftarrow i + \BWT.\rank(v, w)$ to have the correct offset in the next step.

\end{enumerate}

\subsection{Construction in VG}

GBWT construction in VG requires a VCF file with \emph{phasing information}. We expect a diploid genome, though some regions may be haploid. Because we need two layers of buffering, we process the VCF file in batches of $s$ samples (default $200$) in order to save memory.

At each variant site and for every haplotype, we determine the path from the previous site to the current site, and extend the buffered path for that haplotype with it. If there is no phasing information at the current site or if we cannot otherwise extend the path $P$, we insert both the path $P$ and its reverse $\greverse{P}$ into the GBWT construction buffer and start a new path. Once the GBWT construction buffer is full (the default size is 100~million), we launch a background thread to insert the batch into the index.

We can \emph{merge} GBWT indexes quickly if the node identifiers do not overlap (e.g.\ indexes for different chromosomes). The records for all nodes $v \in V$ can be reused in the merged index. In the endmarker record $\$$, we merge the local alphabets and concatenate the record bodies in some order. When we interleave the global haplotype identifier structures, we have to update the identifiers according to the order we used in the endmarker.

\section{Haplotype-aware graph simplification}

VG uses \emph{pruning heuristics} to simplify graphs for \kmer{k} indexing. First we remove edges on \kmer{k}s that make too many edge choices (e.g.\ more than $3$~choices in a \kmer{24}). Edges on unary paths are not deleted, as there is no choice in taking them. Then we delete \emph{connected components} with too little sequence (e.g.\ less than $33$~bases). Finally, if the graph contains reference paths, we may restore them to the pruned graph.

Heuristic pruning often breaks paths taken by known haplotypes. This may cause errors in read mapping, if we cannot find candidate positions for a read in the correct graph region. On the other hand, indexing too many recombinations may increase the number of false positives. Hence we would like to prune recombinations while leaving the haplotypes intact.

\begin{figure}
\includegraphics[width=\textwidth]{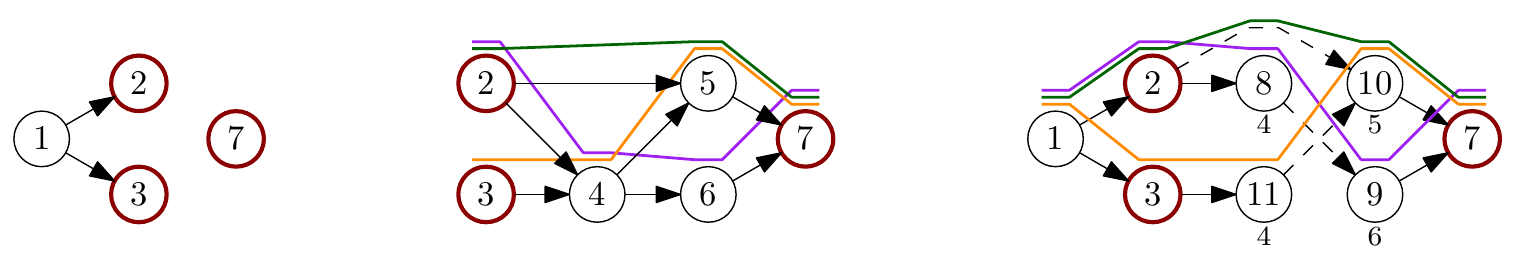}
\caption{Unfolding the paths in the graph in Figure~\ref{fig:gbwt-example}. Border nodes have been highlighted. Left: The graph after removing nodes $4$, $5$, and $6$. Center: Complement graph. The maximal paths are $(2, 4 \mid 6, 7)$, $(2 \mid 5, 7)$, and $(3, 4 \mid 5, 7)$, with the bar splitting a path into a prefix and a suffix. Right: Unfolded graph. Dashed edges cross from prefixes to suffixes. Duplicated nodes have the original ids below the node.}\label{fig:unfolding-example}
\end{figure}

We describe an algorithm that \emph{unfolds} the haplotype paths in pruned regions, \emph{duplicating} nodes when necessary. Our algorithm works with any pruning algorithm that removes nodes from the graph. See Figure~\ref{fig:unfolding-example} for an example. We work with bidirected VG graphs, unless otherwise noted. Reference paths can also be unfolded with a similar algorithm.

Let $G_{i} = (V_{i}, E_{i})$ be the graph induced by GBWT paths and $G_{p} = (V_{p}, E_{p})$ be a pruned graph. We build a \emph{complement graph} induced by edges $E_{i} \setminus E_{p}$ and consider each connected component $G_{c} = (V_{c}, E_{c})$ in it separately. The set $V_{b} = V_{c} \cap V_{p}$ is the \emph{border} of the component, as the nodes exist both in the component and in the pruned graph. Nodes in the set $V_{c} \setminus V_{b}$ are \emph{internal nodes}.

Each connected component $G_{c}$ represents a graph region that was removed from the original graph. We build an unfolded component consisting of the paths in $G_{c}$ supported by GBWT paths and insert it into the pruned graph $G_{p}$. We achieve this by duplicating the internal nodes that would otherwise cause recombinations.

In order to build the unfolded component, we must find all \emph{maximal} paths $P$ of length $\abs{P} \ge 2$ supported by GBWT paths in the component. A path starting from a border node is maximal if it reaches the border again or cannot be extended any further. GBWT paths consisting entirely of internal nodes of the component are also maximal.

Let $v$ be a GBWT node and $vg(v) \in V_{c}$ the corresponding VG node. If $vg(v)$ is a border node, we create a \emph{search state} $(v, \find(v))$ consisting of a pattern and a range. For internal nodes, we create state $(v, \find(\$v))$. Then, for each search state $(X, [sp, ep])$, with $x = X[\abs{X}-1]$:
\begin{enumerate}

\item If $\abs{X} \ge 2$ and the last node $vg(x)$ is a border node, we stop the search for this state. If $vg(X[0])$ is also a border node, $X$ is a maximal path, and we output it.

\item We try to extend the search with all GBWT nodes $v$ corresponding to the successors $u \in V_{c}$ of $vg(x)$, taking the orientation of $v$ from the VG edge. If $[sp', ep'] = [\LF((x, sp), v), \LF((x, ep + 1), v) - 1] \ne \emptyset$, we create a new state $(Xv, [sp', ep'])$.

\item If no extension was successful and $\abs{X} \ge 2$, path $X$ is maximal, and we output it.

\end{enumerate}

Let $P$ be a maximal path we output. If $P$ is not a border-to-border path, we try to extend the lexicographically smaller of $P$ and $\greverse{P}$ with reference paths, replacing $P$ with the extended path. To avoid having the same path in both orientations, we replace each path $P$ with the smaller of $P$ and $\greverse{P}$.

We could create new duplicates of all internal nodes on $P$ and insert the path into $G_{p}$, but this would create too much nondeterminism for GCSA2.\footnote{If VG node $v$ has predecessors $u$ and $u'$ with identical labels, \kmer{k}s starting from $u$ and $u'$ and passing through $v$ cannot be distinguished. GCSA2 construction has to extend these \kmer{k}s until the order of the index (e.g.\ $k = 256$), which may increase the size of the temporary files significantly.} Instead, we split each path into a prefix and a suffix of equal length and build a trie of the prefixes and a trie of the reverse suffixes. Every edge in the tries becomes a node in the unfolded component.

Let $v$ be the label of a trie edge starting from the root. If $vg(v)$ is a border node, it already exists in $G_{p}$. Otherwise we add a new duplicate of $vg(v)$. Now let $v$ and $v'$ be the labels of two successive trie edges, and let $u$ be the VG node we used for $v$. We create a new duplicate $u'$ of $vg(v')$ and add node $u'$ to $G_{p}$. We also add edge $(u, u')$ or $(u', u)$, depending on whether we are in a prefix or a suffix. Finally, if we used VG node $u$ for the end of a prefix and VG node $u'$ for the start of the corresponding suffix, we add edge $(u, u')$ to $G_{p}$.

After we have handled all components, the simplified graph $G_{p}$ contains all GBWT paths. The GCSA2 index of $G_{p}$ contains all \kmer{k}s (e.g.\ \kmer{256}s) in the haplotypes. This allows us to prune the graph more aggressively, removing more \kmer{k}s corresponding to recombinations. In order to map reads to the original graph $G = (V, E)$ instead of the simplified graph $G_{p}$, we replace the node identifiers $v \in V_{p}$ in the GCSA2 index with the original identifiers $v' \in V$.

\section{Experiments}

We have implemented GBWT in C++ using the SDSL library \cite{Gog2014b}. The following experiments were done using VG~v1.7.0 with prerelease versions of GBWT~v0.4 and GCSA2~v1.2. All code was compiled using GCC~5.4. We used a single Amazon~EC2 i3.8xlarge instance with 16~physical (32~logical) cores of an Intel Xeon E5~2686~v4 and 244~GiB\footnote{Sizes measured in MiB, GiB, and TiB are based on 1024-byte kibibytes. Sizes measured in MB, GB, and TB are based on 1000-byte kilobytes.} of memory. The system was running Ubuntu~16.04 with Linux kernel~4.4.0. The temporary files in GCSA2 construction were stored on a local RAID~0 volume consisting of four 1.9~TB SSDs.

\subsection{GBWT construction}

We built VG graphs from the GRCh37 human reference genome and the \emph{1000~Genomes Project} (1000GP) final phase data \cite{1000GP2015}. The VCF files had phasings for 2504~humans over approximately 80~million variants. VG transformed the phasings into 29.3~million paths of total length 1.62~trillion in a graph with 493~million nodes (including the reverse paths).

GBWT construction is space-efficient and uses two threads. We first built separate GBWTs for each chromosome, running 12~jobs in parallel. The jobs were ordered $X, 1, \dotsc, 22, Y$, as large chromosomes take longer to finish. Total construction time was 29.0~hours. The longest job was 27.1~hours for chromosome~2. The bottleneck was generating haplotype paths, as the insertion threads were running less than half of the total time.

Merging the GBWTs into a single index took less than 9~minutes. The merged index took 14.6~GiB, out of which 7.4~GiB was for the GBWT itself and 7.2~GiB for the haplotype identifiers ($d = 1024$). The dynamic GBWT was roughly 10x larger. Its exact size is not well-defined due to a large number of memory allocations and unused space in the arrays.

All the assumptions in Section~\ref{sect:records} were valid for our dataset:
\begin{enumerate}

\item The average outdegree of a record is $1.34$ and the maximum is $13$, excluding the endmarker.

\item As the graphs built from a VCF file are acyclic, no haplotype can visit the same node twice.

\item The GBWT takes 0.04~bits per character, excluding the haplotype identifiers.

\item VG construction avoids leaving gaps between node identifiers.

\item The VG graphs built from a VCF file are almost in topological order.

\end{enumerate}

\subsection{GBWT benchmarks}

For various pattern lengths $\abs{X}$ from $2$ to $50$, we extracted 100,000 patterns from the whole-genome index and used them for $\find()$ queries. The average query times in the compressed GBWT start from 460~ns/character with $\abs{X} = 2$ and go down to 300~ns/character with $\abs{X} = 50$ due to memory locality. This is somewhat slower than in FM-indexes over small alphabets \cite{Siren2017}. For the dynamic GBWT, query times were 130~ns/character with $\abs{X} = 2$ and 80~ns/character with $\abs{X} = 50$, or 2--3 times faster than in ordinary FM-indexes.

The $\locate()$ performance suffers from the long distance $d = 1024$ between stored identifiers. We extracted 20,000 patterns of length $20$ from the index and used the ranges returned by $\find()$ queries for $\locate()$ benchmarks. The total length of the ranges was 69.1~million. The average query times in the compressed GBWT were 110~\textmu{}s/position (direct algorithm) and 14~\textmu{}s/position (fast algorithm). For the dynamic index, the times were 19~\textmu{}s/position (direct) and 11~\textmu{}s/position (fast). FM-indexes for non-repetitive text typically use $d = 16$ or $d = 32$ and take a few microseconds to locate each position.

We also extracted 100,000 paths of total length 5.50~billion from the index. The average time per character was 1,800~ns in the compressed index and 410~ns in the dynamic index. This is several times slower than in $\find()$ queries: $\find()$ uses $\LF((v, i), c)$ instead of the slower $\LF(v, i)$, while $\extract()$ queries start from the very large record for the endmarker $\$$. While the direct $\locate()$ algorithm also uses $\LF(v, i)$, it benefits from memory locality, as it traverses the same graph region for each position in the query range.

\subsection{Haplotype-aware graphs}

The typical pruning parameters for a \kmer{128} GCSA2 index are $4$ edge choices in a \kmer{16}. When building \kmer{256} indexes, we need to prune more aggressively: $3$ edge choices in a \kmer{24}. This removes more \kmer{k}s corresponding to both haplotypes and their recombinations. In the following, graph \pruned{k} has been pruned with the parameters for a \kmer{k} index, and the reference paths have been restored afterwards. Similarly, \unfolded{k} is a graph, where the haplotype paths and reference paths have been unfolded after pruning.

\begin{table}[t!]
\caption{GCSA2 indexes for simplified 1000GP graphs. Index size in GiB and construction time in hours. Number of distinct \kmer{64}s (in billions) shared with \pruned{256}, additional haplotype \kmer{64}s, and additional recombination \kmer{64}s.}\label{table:gcsa2}
\begin{center}
\begin{tabular}{c|cc|ccc}
\hline
 & \multicolumn{2}{c|}{\textbf{GCSA2 index}} & \multicolumn{3}{c}{\textbf{\kmer{\mathbf{64}}s}} \\
\textbf{Graph} & \textbf{Size} & \textbf{Construction} & \textbf{Shared} & \textbf{Haplotype} & \textbf{Recombination} \\
\hline
\pruned{128}   & 35.4~GiB & 25.4~h & 27.0~G & 3.11~G & 11.4~G \\
\pruned{256}   & 29.3~GiB & 25.5~h & 27.0~G &     -- & -- \\
\unfolded{256} & 33.7~GiB & 28.9~h & 27.0~G & 3.46~G & -- \\
\hline
\end{tabular}
\end{center}
\end{table}

We created simplified whole-genome graphs \pruned{128}, \pruned{256}, and \unfolded{256}. The simplification took 3--4~hours for \unfolded{256} and slightly less for the other graphs. We then built GCSA2 indexes for both orientations of the simplified graphs. The results can be seen in Table~\ref{table:gcsa2}. The index for \pruned{256} is a few GiB smaller than the others, while the index for \unfolded{256} takes a few hours longer to build. Graph \pruned{128} contains 90~\% of the haplotype \kmer{k}s missing from \pruned{256} but included in \unfolded{256}. (Some haplotype \kmer{k}s may be recombinations crossing between simple and unfolded regions.) It also contains a large number of additional recombination \kmer{k}s not present in \unfolded{256}.

\section{Discussion}

We have developed GBWT, a scalable implementation of the graph extension of the PBWT. The earlier implementation used 9.3~hours and 278~GiB of memory to index the 1000GP chromosome~22 using a single thread \cite{Novak2017a}. In comparison, our implementation takes 4.1~hours and less than 10~GiB of memory using two threads. We also reduced the final index size from 321~MiB to 110~MiB (without haplotype identifiers). By running multiple jobs in parallel, we were able to build a whole-genome GBWT in less than 30~hours on a single system.

Contemporary sequencing projects are sequencing in excess of 100,000 diploid genomes. We intend to scale GBWT to allow working with such large collections, providing a compressed, indexed and searchable representation that should fit into the memory of a single server. Potential applications in genome inference and imputation, as well as for powering population genomic queries, are myriad. The main bottleneck here is construction time. We currently parse the VCF file and find the paths between variant sites once for every 200~samples, which takes the bulk of the time. By parsing the file once and storing the information in a directly usable format, we should be able to double the construction speed.

Storing the haplotype identifiers for $\locate()$ queries is another bottleneck. With 5,000~haplotypes, the identifiers use roughly as much space as the GBWT itself. If we increase the number of haplotypes to 50,000, GBWT size should not increase too much, while the identifiers will take 10x more space. We can save space by increasing the distance between stored identifiers, at the expense of increased query times. There is a theoretical proposal for supporting fast $\locate()$ queries in space proportional to the size of the run-length encoded BWT \cite{Gagie2018}. Building the proposed structure for large text collections is still an open problem.

We used the haplotype information in GBWT to simplify VG graphs for \kmer{k} indexing. This allowed us to prune the \kmer{k}s corresponding to recombinations more aggressively, while still having all \kmer{k}s from the haplotypes in the index. CHOP, the other haplotype-aware graph indexing approach, can only use short-range haplotype information in read mapping. Because VG graphs are connected, we can use the long-range information in the GBWT for mapping long reads and paired-end reads. We will investigate this in a subsequent paper.

\bibliography{gbwt-paper}

\begin{thebibliography}{10}

\bibitem{Bauer2013}
Markus~J. Bauer, Anthony~J. Cox, and Giovanna Rosone.
\newblock Lightweight algorithms for constructing and inverting the {BWT} of
  string collections.
\newblock {\em Theoretical Computer Science}, 483:134--148, 2013.
\newblock \href {http://dx.doi.org/10.1016/j.tcs.2012.02.002}
  {\path{doi:10.1016/j.tcs.2012.02.002}}.

\bibitem{Burrows1994}
Michael Burrows and David~J. Wheeler.
\newblock A block sorting lossless data compression algorithm.
\newblock Technical Report 124, Digital Equipment Corporation, 1994.
\newblock URL:
  \url{http://www.hpl.hp.com/techreports/Compaq-DEC/SRC-RR-124.html}.

\bibitem{Chan2007}
Ho-Leung Chan et~al.
\newblock Compressed indexes for dynamic text collections.
\newblock {\em ACM Transactions on Algorithms}, 3(2):21, 2007.
\newblock \href {http://dx.doi.org/10.1145/1240233.1240244}
  {\path{doi:10.1145/1240233.1240244}}.

\bibitem{Durbin2014}
Richard Durbin.
\newblock Efficient haplotype matching and storage using the {P}ositional
  {B}urrows--{W}heeler transform ({PBWT}).
\newblock {\em Bioinformatics}, 30(9):1266--1272, 2014.
\newblock \href {http://dx.doi.org/10.1093/bioinformatics/btu014}
  {\path{doi:10.1093/bioinformatics/btu014}}.

\bibitem{Eggertsson2017}
Hannes~P. Eggertsson et~al.
\newblock Graphtyper enables population-scale genotyping using pangenome
  graphs.
\newblock {\em Nature Genetics}, 49:1654--1660, 2017.
\newblock \href {http://dx.doi.org/10.1038/ng.3964}
  {\path{doi:10.1038/ng.3964}}.

\bibitem{Ferragina2005a}
Paolo Ferragina and Giovanni Manzini.
\newblock Indexing compressed text.
\newblock {\em Journal of the ACM}, 52(4):552--581, 2005.
\newblock \href {http://dx.doi.org/10.1145/1082036.1082039}
  {\path{doi:10.1145/1082036.1082039}}.

\bibitem{Gagie2017a}
Travis Gagie, Giovanni Manzini, and Jouni Sirén.
\newblock Wheeler graphs: A framework for {BWT}-based data structures.
\newblock {\em Theoretical Computer Science}, 698:67--78, 2017.
\newblock \href {http://dx.doi.org/10.1016/j.tcs.2017.06.016}
  {\path{doi:10.1016/j.tcs.2017.06.016}}.

\bibitem{Gagie2018}
Travis Gagie, Gonzalo Navarro, and Nicola Prezza.
\newblock Optimal-time text indexing in {BWT}-runs bounded space.
\newblock In {\em Proc.\ ALENEX 2018}, pages 1459--1477. SIAM, 2018.
\newblock \href {http://dx.doi.org/10.1137/1.9781611975031.96}
  {\path{doi:10.1137/1.9781611975031.96}}.

\bibitem{Garrison2017}
Erik Garrison et~al.
\newblock Sequence variation aware genome references and read mapping with the
  variation graph toolkit.
\newblock bioRxiv, 2017.
\newblock \href {http://dx.doi.org/10.1101/234856} {\path{doi:10.1101/234856}}.

\bibitem{Gog2014b}
Simon Gog et~al.
\newblock From theory to practice: Plug and play with succinct data structures.
\newblock In {\em Proc.\ SEA 2014}, volume 8504 of {\em LNCS}, pages 326--337.
  Springer, 2014.
\newblock \href {http://dx.doi.org/10.1007/978-3-319-07959-2_28}
  {\path{doi:10.1007/978-3-319-07959-2_28}}.

\bibitem{Hon2007}
Wing-Kai Hon et~al.
\newblock A space and time efficient algorithm for constructing compressed
  suffix arrays.
\newblock {\em Algorithmica}, 48(1):23--36, 2007.
\newblock \href {http://dx.doi.org/10.1007/s00453-006-1228-8}
  {\path{doi:10.1007/s00453-006-1228-8}}.

\bibitem{Huang2013}
Lin Huang, Victoria Popic, and Serafim Batzoglou.
\newblock Short read alignment with populations of genomes.
\newblock {\em Bioinformatics}, 29(13):i361--i370, 2013.
\newblock \href {http://dx.doi.org/10.1093/bioinformatics/btt215}
  {\path{doi:10.1093/bioinformatics/btt215}}.

\bibitem{Huang2010}
Songbo Huang et~al.
\newblock Indexing similar {DNA} sequences.
\newblock In {\em Proc.\ AAIM 2010}, volume 6124 of {\em LNCS}, pages 180--190.
  Springer, 2010.
\newblock \href {http://dx.doi.org/10.1007/978-3-642-14355-7_19}
  {\path{doi:10.1007/978-3-642-14355-7_19}}.

\bibitem{Li2012}
Heng Li.
\newblock Exploring single-sample {SNP} and {INDEL} calling with whole-genome
  de novo assembly.
\newblock {\em Bioinformatics}, 28(14):1838--1844, 2012.
\newblock \href {http://dx.doi.org/10.1093/bioinformatics/bts280}
  {\path{doi:10.1093/bioinformatics/bts280}}.

\bibitem{Li2014a}
Heng Li.
\newblock Fast construction of {FM}-index for long sequence reads.
\newblock {\em Bioinformatics}, 30(22):3274--3275, 2014.
\newblock \href {http://dx.doi.org/10.1093/bioinformatics/btu541}
  {\path{doi:10.1093/bioinformatics/btu541}}.

\bibitem{Maciuca2016}
Sorina Maciuca et~al.
\newblock A natural encoding of genetic variation in a {B}urrows-{W}heeler
  transform to enable mapping and genome inference.
\newblock In {\em Proc.\ WABI 2016}, volume 9838 of {\em LNCS}, pages 222--233.
  Springer, 2016.
\newblock \href {http://dx.doi.org/10.1007/978-3-319-43681-4_18}
  {\path{doi:10.1007/978-3-319-43681-4_18}}.

\bibitem{Mokveld2018}
Tom~O. Mokveld et~al.
\newblock {CHOP}: Haplotype-aware path indexing in population graphs.
\newblock bioRxiv, 2018.
\newblock \href {http://dx.doi.org/10.1101/305268} {\path{doi:10.1101/305268}}.

\bibitem{Maekinen2010}
Veli Mäkinen et~al.
\newblock Storage and retrieval of highly repetitive sequence collections.
\newblock {\em Journal of Computational Biology}, 17(3):281--308, 2010.
\newblock \href {http://dx.doi.org/10.1089/cmb.2009.0169}
  {\path{doi:10.1089/cmb.2009.0169}}.

\bibitem{Na2016}
Joong~Chae Na et~al.
\newblock {FM}-index of alignment: A compressed index for similar strings.
\newblock {\em Theoretical Computer Science}, 638:159--170, 2016.
\newblock \href {http://dx.doi.org/10.1016/j.tcs.2015.08.008}
  {\path{doi:10.1016/j.tcs.2015.08.008}}.

\bibitem{Na2018}
Joong~Chae Na et~al.
\newblock {FM}-index of alignment with gaps.
\newblock {\em Theoretical Computer Science}, 710(148-157), 2018.
\newblock \href {http://dx.doi.org/10.1016/j.tcs.2017.02.020}
  {\path{doi:10.1016/j.tcs.2017.02.020}}.

\bibitem{Novak2017a}
Adam Novak, Erik Garrison, and Benedict Paten.
\newblock A graph extension of the positional {B}urrows–{W}heeler transform
  and its applications.
\newblock {\em Algorithms for Molecular Biology}, 12:18, 2017.
\newblock \href {http://dx.doi.org/10.1186/s13015-017-0109-9}
  {\path{doi:10.1186/s13015-017-0109-9}}.

\bibitem{Okanohara2007}
Daisuke Okanohara and Kunihiko Sadakane.
\newblock Practical entropy-compressed rank/select dictionary.
\newblock In {\em Proc.\ ALENEX 2007}, pages 60--70. SIAM, 2007.
\newblock \href {http://dx.doi.org/10.1137/1.9781611972870.6}
  {\path{doi:10.1137/1.9781611972870.6}}.

\bibitem{Rakocevic2017}
Goran Rakocevic et~al.
\newblock Fast and accurate genomic analyses using genome graphs.
\newblock bioRxiv, 2017.
\newblock \href {http://dx.doi.org/10.1101/194530} {\path{doi:10.1101/194530}}.

\bibitem{Schneeberger2009}
Korbinian Schneeberger et~al.
\newblock Simultaneous alignment of short reads against multiple genomes.
\newblock {\em Genome Biology}, 10(9):R98, 2009.
\newblock \href {http://dx.doi.org/10.1186/gb-2009-10-9-r98}
  {\path{doi:10.1186/gb-2009-10-9-r98}}.

\bibitem{Siren2017}
Jouni Sirén.
\newblock Indexing variation graphs.
\newblock In {\em Proc.\ ALENEX 2017}, pages 13--27. SIAM, 2017.
\newblock \href {http://dx.doi.org/10.1137/1.9781611974768.2}
  {\path{doi:10.1137/1.9781611974768.2}}.

\bibitem{Siren2014}
Jouni Sirén, Niko Välimäki, and Veli Mäkinen.
\newblock Indexing graphs for path queries with applications in genome
  research.
\newblock {\em IEEE/ACM Transactions on Computational Biology and
  Bioinformatics}, 11(2):375--388, 2014.
\newblock \href {http://dx.doi.org/10.1109/TCBB.2013.2297101}
  {\path{doi:10.1109/TCBB.2013.2297101}}.

\bibitem{1000GP2015}
{The 1000 Genomes Project Consortium}.
\newblock A global reference for human genetic variation.
\newblock {\em Nature}, 526:68--64, 2015.
\newblock \href {http://dx.doi.org/10.1038/nature15393}
  {\path{doi:10.1038/nature15393}}.

\bibitem{Marschall2018}
{The Computational Pan-Genomics Consortium}.
\newblock Computational pan-genomics: status, promises and challenges.
\newblock {\em Briefings in Bioinformatics}, 19(1):118--135, 2018.
\newblock \href {http://dx.doi.org/10.1093/bib/bbw089}
  {\path{doi:10.1093/bib/bbw089}}.

\end{thebibliography}

\end{document}